# Equity, diversity, and inclusion in sports analytics

Craig Fernandes[1], Jason D. Vescovi[2], Richard Norman[3], Cheri L. Bradish[3], Nathan Taback[4], Timothy C.Y. Chan[1]

[1]Department of Mechanical and Industrial Engineering, University of Toronto
[2]USA Lacrosse's Center for Sport Science
[3]Ted Rogers School of Management, Ryerson University
[4]Department of Statistical Sciences, University of Toronto

This paper presents a landmark study of equity, diversity and inclusion (EDI) in the field of sports analytics. We developed a survey that examined personal and job-related demographics, as well as individual perceptions and experiences about EDI in the workplace. We sent the survey to individuals in the five major North American professional leagues, representatives from the Olympic and Paralympic Committees in Canada and the U.S., the NCAA Division I programs, companies in sports tech/analytics, and university research groups. Our findings indicate the presence of a clear dominant group in sports analytics identifying as: young (72.0%), White (69.5%), heterosexual (89.7%) and male (82.0%). Within professional sports, males in management positions earned roughly $30,000 (27%) more on average compared to females. A smaller but equally alarming pay gap of $17,000 (14%) was found between White and non-White management personnel. Of concern, females were nearly five times as likely to experience discrimination and twice as likely to have considered leaving their job due to isolation or feeling unwelcome. While they had similar levels of agreement regarding fair processes for rewards and compensation, females "strongly agreed" less often than males regarding equitable support, equitable workload, having a voice, and being taken seriously. Over one third (36.3%) of females indicated that they "strongly agreed" that they must work harder than others to be valued equally, compared to 9.8% of males. We conclude the paper with concrete recommendations that could be considered to create a more equitable, diverse and inclusive environment for individuals working within the sports analytics sector.

## 1. Introduction

The field of sports analytics has grown tremendously over the last 20 years. This rise has been accompanied by significant growth in analytics-focused personnel being hired into the sports industry, such as by sports teams and sport technology companies. Many analytics-based industries, which recruit heavily from the fields of science, technology, engineering, and math (STEM), have noted a persistent lack of diversity within their workforce (Funk and Parker 2018; Built In 2021). In professional and amateur sports, there have been many well-documented examples of discrimination against and exclusion of athletes and coaches who belong to groups considered marginalized in society (Kahn 1991; Ayala 2020; Cunningham, Wicker and Walker 2021; Mervosh and Caron 2020). Preliminary findings suggests that the intersection of these two fields – sports analytics – suffers from the same issues (Benbow 2021). However, a formal study has yet to be conducted. According to our knowledge, this paper is the first to evaluate the state of equity, diversity, and inclusion (EDI) in sports analytics.

Before proceeding, we first define the various components of EDI (Built In 2021; University of Toronto 2019). Equity refers to the practice of fair and impartial access across the organization. Equitable processes ensure that each individual, irrespective of background, has equal opportunities. Diversity refers to the presence of individuals whose characteristics span all dimensions of race, sex, gender, sexual orientation, age, religious belief, or physical ability in accordance with the overall population. Inclusion refers to feeling a sense of belonging and the ability of individuals to be their authentic selves.

The value of organizations embracing EDI has been well-documented. For example, studies have shown that individuals reported higher levels of well-being and workplace satisfaction when working for a company that valued and implemented EDI initiatives (Findler, Wind and Barak 2007; Wronski 2021). Accordingly, a recent survey of U.S. workers showed that the majority of employees (78%) believe it is important for them to "work at an organization that prioritizes diversity and inclusion" (Wronski 2021). This proportion was even higher for marginalized and racialized communities (85%-88%) likely due to the disproportionate negative consequences these communities encounter in organizations that do not value EDI. For example, a 2017 meta-analysis study (Quillian et al. 2017) showed that Black and Latinx Americans face more hiring discrimination than their White colleagues, a phenomenon that EDI best practices seeks to address. Within Canada, Oud (2018) showed that Canadian university librarians with disabilities had lower levels of satisfaction regarding their work-related stress levels, the amount of support they receive (from supervisors and colleagues) and the flexibility they have in their schedule, compared to librarians without disabilities. There is also a large body of research confirming the organizational benefits of embracing EDI. For example, companies with diverse workforces have been shown to have increased creativity, productivity, knowledge-sharing, and financial performance (Richard, Triana and Li 2021; Saxena 2014; Dixon-Fyle et al. 2020). Given this background, there is a clear case for developing a thorough understanding of the state of equity, diversity, and inclusion in the field of sports analytics.

In this study, we surveyed individuals working in sports analytics to measure personal and job-related demographic variables, as well as their perceptions and experiences with respect to EDI in the workplace. Our primary findings are:

1. **A clear dominant group.** Respondents predominantly self-reported being young (72.0%), White (69.5%), heterosexual (89.7%) and male (82.0%). Within professional sports, this imbalance seems to grow as age increases and persists across job levels.
2. **Pay disparity.** Within professional sports, males in management positions earned roughly $30,000 (27%) more on average compared to females. A smaller pay gap of $17,000 (14%) was found between White and non-White management personnel. For non-management personnel, smaller pay gaps persisted.
3. **Greater barriers for females.** Females had lower workplace satisfaction than males. For example, females were nearly five times as likely to experience discrimination and twice as likely to have considered leaving their job due to isolation or feeling unwelcome. While they had similar levels of agreement regarding fair processes for rewards and compensation, females "strongly agreed" less often than males regarding equitable support, equitable workload, having a voice, and being taken seriously. Over a third (36.3%) of females "strongly agreed" that they must work harder than others to be valued equally, compared to 9.8% of males.

We conclude the paper with concrete recommendations that could lead to a more equitable, diverse and inclusive environment for individuals working within the sports analytics sector.

## 2. Survey Methods

The survey was developed by following established best practices for collecting race- and EDI-based data (Government of Ontario 2021). The survey consisted of 27 questions that contained multiple choice and Likert scale response options. Information collected included personal demographic information such as age, race, relationship status, sex, and gender, and job-related information such as their current organization type, department type, salary, and tenure at the organization. Respondents were also asked about their experiences and perceptions with respect to inclusion, such as how often they experienced discrimination and whether they feel valued by their employer. Finally, we gathered information about whether the COVID-19 pandemic has influenced their responses. The complete survey is presented in Appendix A.

The survey was administered via Microsoft Forms and was made available on the webpage http://sportsanalytics.utoronto.ca/edi-survey between April 12 – June 4, 2021. The survey was anonymous, and respondents were aware of this fact when filling out the survey. Efforts were made to disseminate the survey as widely as possible. We sent the survey to sports analytics and other personnel working in the major North American professional leagues (MLB, MLS, NBA, NFL, NHL), National Sport Organizations (i.e., Olympic and Paralympic) in Canada and the U.S., NCAA Division I programs, sports technology and sports analytics companies, and university research groups working in sports analytics. We identified appropriate individuals based on their department title or job title (i.e., titles that included keywords such as "analytics", "data", "high performance", "statistics", etc.). Contact information for these individuals came from publicly available sources (e.g., an organization's website, LinkedIn), via the study team's personal contact lists, and through referrals from the individuals we contacted. In some cases, an individual working at a professional league was the primary contact, and that individual helped forward the survey to individuals from different teams, rather than our study team contacting the teams directly. In total, the study team sent the survey to 420 individuals via email, which included individuals at 27 professional sport organizations; 90 National sport organizations; 264 NCAA universities; 16 sports technology and sports analytics companies; and 23 universities with sport analytics groups.

The MIT Sloan Sports Analytics Conference (SSAC) Twitter account tweeted the survey to its ~34.6K followers – this channel had the largest reach by far. The tweet was shared on May 24, 2021, and we received 20 subsequent responses, but are not able to confirm if these individuals accessed the survey due to the tweet. Team members also publicized the survey at events they attended such as virtual panel sessions during the survey period. We strongly encouraged anybody who received the survey to share it further among their own networks. We were not able to track how many people were forwarded the survey from other parties.

We avoided collecting any potentially identifying information (e.g., the respondent's organization) to maintain the anonymity of the survey. As a result, we cannot confirm which teams or leagues filled out the survey. For example, while we can identify respondents within professional sports, we cannot comment on how many of them came from the NHL versus the NBA, and within each league, how many came from a particular team. Moreover, we cannot identify the non-respondents of our survey sample or characteristics about these non-respondents.

We received a total of 260 survey responses. Twenty-seven responses were removed because they answered "no" to the first two questions (see Appendix), which indicated that the respondent was not involved in sports analytics. The remaining 233 responses were included in the analysis presented in the following section. We conducted both univariate and cross-tabulated analysis of personal and job-related demographics, as well as experiences with EDI in the workplace. We performed limited statistical testing (e.g., hypothesis testing) due to our limited sample size and lack of data in particular stratifications.

## 3. Results

This section presents our analysis of the 233 eligible responses from our survey. We provide deeper interpretations of these results in Section 4.

### 3.1. Analysis of demographic variables

Table 1 summarizes personal features. Among all respondents, 72.0% were younger than 35 years old, 82.0% identified as male, 89.7% identified as heterosexual and 69.5% identified as White (excluding mixed race). One third (33.0%) of respondents possessed all these features, representing the largest cohort among all combinations of these variables. In addition, 5.2% of respondents reported having at least one disability (e.g., mental-health, vision, hearing, mobility, etc.). Nearly three quarters of respondents had either a Bachelor's or Master's degree (75.6%) as their highest degree, and nearly 20% of respondents had a Doctoral degree. Respondents hailed from many different disciplines. Statistics, mathematics, engineering and business management were the primary fields of study for 60.9% of respondents. Sports-related fields such as kinesiology, sport science and sport medicine were less prevalent (10.3%). Fields of study in the "Other" category largely consisted of social sciences and humanities.

*Table 1. Personal-related demographics of the 233 included respondents*

| Variable | n (%) | Variable | n (%) |
|---|---|---|---|
| **Age group, yr** | | **Relationship Status** | |
| <19 | 1 (0.4) | Married/partnership | 94 (40.3) |
| 20-24 | 31 (13.3) | Single | 75 (32.2) |
| 25-29 | 73 (31.3) | Committed relationship | 58 (24.9) |
| 30-34 | 63 (27.0) | P.N.T.A | 5 (2.1) |
| 35-39 | 28 (12.0) | Divorced | 1 (0.4) |
| 40-44 | 15 (6.4) | **Gender** | |
| 45-49 | 11 (4.7) | Man/Primarily masculine | 186 (79.8) |
| 50+ | 10 (4.3) | Woman/Primarily feminine | 35 (15.0) |
| Missing | 1 (0.4) | P.N.T.A | 8 (3.4) |
| **Sex** | | Neither, trans, or other gender minority | 4 (1.7) |
| Male | 191 (82.0) | **Highest Degree Earned** | |
| Female | 35 (15.0) | Doctoral degree | 46 (19.7) |
| P.N.T.A[1] | 7 (3.0) | Master's degree | 88 (37.8) |
| **Sexual Orientation** | | Law degree | 2 (0.9) |
| Heterosexual | 209 (89.7) | Bachelor's degree (4yr) | 88 (37.8) |
| LGBTQP | 16 (6.9) | Associates degree (2yr) | 2 (0.9) |
| P.N.T.A | 8 (3.4) | High school | 6 (2.6) |
| **Race** | | Missing | 1 (0.4) |
| White | 162 (69.5) | **Field of Study** | |
| Mixed Race | 20 (8.6) | Stats/Math | 80 (34.3) |
| E/SE Asian | 17 (7.3) | Engineering | 35 (15.0) |
| S Asian | 14 (6.0) | Business management | 27 (11.6) |
| P.N.T.A | 6 (2.6) | Other | 25 (10.7) |
| Latinx | 5 (2.1) | Kinesiology/Sport science | 20 (8.6) |
| Black | 4 (1.7) | Computer science | 15 (6.4) |
| Middle Eastern | 3 (1.3) | Economics | 11 (4.7) |
| Missing | 2 (0.9) | Missing | 8 (3.7) |
| **Disability** | | Physics | 5 (2.1) |
| None | 216 (92.7) | Sports management | 3 (1.3) |
| 1+ | 12 (5.2) | Marketing | 3 (1.3) |
| P.N.T.A | 5 (2.1) | Sport medicine | 1 (0.4) |

Table 2 summarizes job-related features of the respondents. The majority of respondents worked for professional sports teams (54.1%). The salary distribution ranged from "volunteer", which we assumed to mean no salary, to >$200K. The distribution was fairly symmetric between $0 to $200K, with 53.2% of all respondents reporting a salary in the range $50K-124.9K. However, the overall distribution had a longer right tail, with 6.9% of respondents reporting a salary above $200K. Over half of the respondents (51.5%) reported being a part of an Analytics department, with over 80% of those individuals working for a

---

[1] P.N.T.A stands for "prefer not to answer".

professional sports team or sport tech company. Kinesiology/sports departments were more prevalent than analytics departments at the amateur and college sport level. The two most common job levels were "Staff" and "Director", together comprising 61.3% of respondents.

*Table 2. Job-related demographics of the 233 included respondents.*

| Variable | n (%) | Variable | n (%) |
|---|---|---|---|
| **Salary** | | **Type of Organization** | |
| >$200K | 16 (6.9) | Professional sport | 126 (54.1) |
| $175-199.9K | 4 (1.7) | Academic/research | 33 (14.2) |
| $150-174.9K | 5 (2.1) | Sport tech company | 29 (12.4) |
| $125-149.9K | 20 (8.6) | Amateur sport | 17 (7.3) |
| $100-124.9K | 44 (18.9) | College sport | 11 (4.7) |
| $75-99.9K | 42 (18.0) | Sport entertainment | 6 (2.6) |
| $50-74.9K | 38 (16.3) | Other | 5 (2.1) |
| $25-49.9K | 25 (10.7) | Media | 4 (1.7) |
| <$25k | 16 (6.9) | Community Sport | 2 (0.9) |
| Volunteer | 8 (3.4) | **Level of Job** | |
| P.N.T.A | 15 (6.4) | Staff | 97 (41.6) |
| **Type of Department** | | Director | 46 (19.7) |
| Analytics | 120 (51.5) | Manager/supervisor | 23 (9.9) |
| Stats/Math | 33 (14.2) | Researcher | 17 (7.3) |
| Sport performance | 22 (9.4) | Faculty | 15 (6.4) |
| Coaching staff | 14 (6.0) | Executive | 13 (5.6) |
| Business operations | 12 (5.2) | Student | 13 (5.6) |
| Engineering | 8 (3.4) | Intern | 6 (2.6) |
| Kin/Sport | 6 (2.6) | Other | 2 (0.9) |
| Computer science | 6 (2.6) | Missing | 1 (0.4) |
| Business development | 3 (1.3) | **Years at current Company** | |
| Tech/digital | 3 (1.3) | <1 yr | 41 (17.6) |
| Other | 2 (0.9) | 1-2 yr | 66 (28.3) |
| Sport medicine | 2 (0.9) | 3-4 yr | 60 (25.8) |
| Marketing | 1 (0.4) | 5-6 yr | 36 (15.5) |
| Missing | 1 (0.4) | >6 yr | 30 (12.9) |

To facilitate interpretation of the results with a more homogenous sample, the following analyses focused on respondents from professional sports (n=126), which was the most prevalent organization type. Figure 1 shows that as age increases, the proportion of female and non-White employees decreases. The majority of respondents (64.3%) were in the 25-34 age range. On the other hand, the 40+ category only had 10 respondents (<1%).

Figure 2 shows that there is a consistent imbalance in sex and race, except at the entry-intern level. Indeed, at each job level, males comprise 85-100% of respondents and Whites comprise 66-85% of respondents. Similar to the previous cross-tabulation with age, some categories had very few observations. For example, four respondents indicated "Intern", six indicated "Researcher", and seven indicated "Executive". Acknowledging that there were very few respondents in the "Intern" category, this was the only job level where non-Whites were in the majority (75%) and where there was parity in sex.

Next, we examined how salary differed by sex and race. Due to small sample sizes for individual job levels, we grouped respondents into either a management (i.e., manager/supervisor, director or executive) or non-management category (Table 3). We took the midpoint of each salary bucket and computed the mean and standard deviation of the salary distribution for the management and non-management categories, stratified by sex and race. For the salary bucket ">$200K", we assumed a value of $250K. We excluded respondents who selected "prefer not to answer". Table 3 shows that males earned more than females at both the management (on average approximately $30K or 27% more) and non-management levels (on average approximately $15K or 23% more). Similarly, White respondents earned more than non-White respondents at the management (on average approximately $17K or 14%) and non-management levels (on average approximately $5K or 7% more).

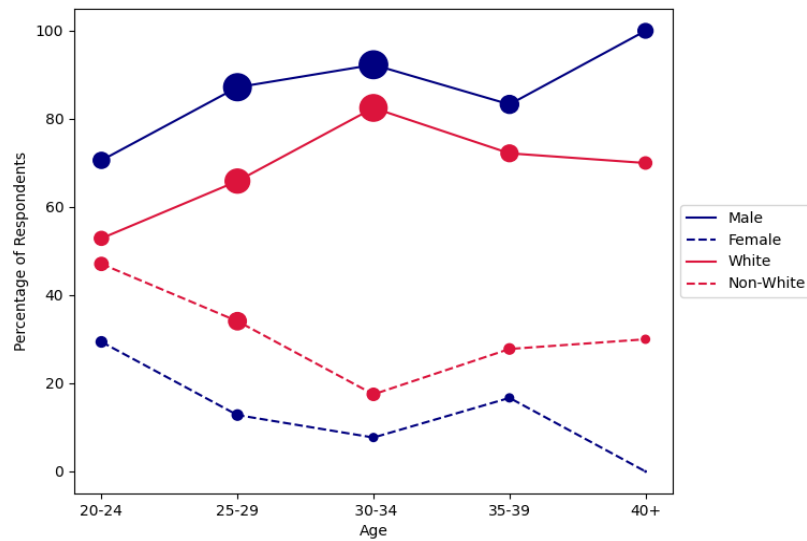

*Figure 1 The percentage of respondents in professional sport, in each age category, broken down by sex and race. The size of each dot represents the relative number of respondents.*

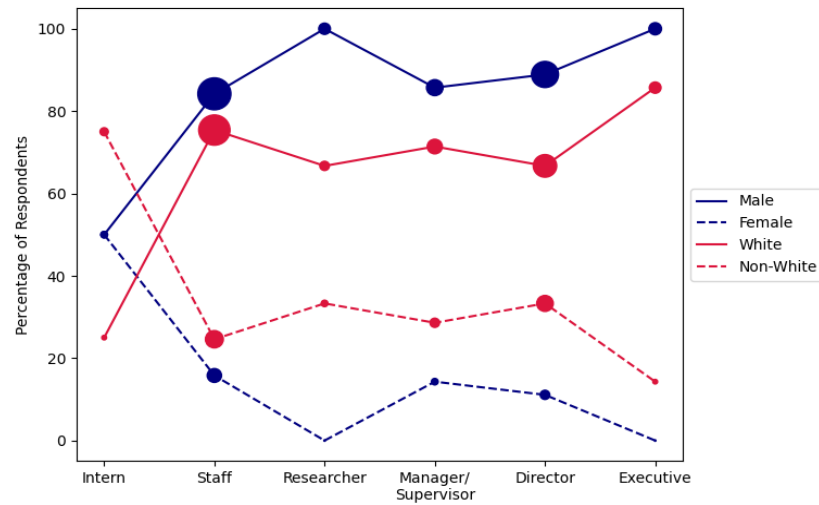

*Figure 2 The percentage of respondents in each job level category, broken down by sex and race. The size of each dot represents the relative number of respondents.*

*Table 3. The salary distribution in professional sport between management and non-management stratified by sex and race.*

|  | Male | Female |  | White | Non-White |
|---|---|---|---|---|---|
| Management (mean ± SD in $,000's) | 138.0 ± 57.2 (n=49) | 108.3 ± 30.3 (n=6) |  | 139.1 ± 57.4 (n=40) | 122.3 ± 48.2 (n=15) |
| Non-management (mean ± SD in $,000's) | 82.7 ± 44.5 (n=58) | 67.5 ± 38.4 (n=10) |  | 81.9 ± 39.8 (n=49) | 76.9 ± 52.3 (n=19) |

The "Management" and "Non-management" categories in Table 3 each consist of three distinct job levels, which may have different salary distributions. To further explore differences in salary by sex and race, we isolated the two most populous job levels within "Management" and "Non-management", namely, "Directors" and "Staff", respectively. Table 4 shows analogous results to Table 3 but focusing on these two specific job levels. Conditioned on the same job level, the salary gap persists between male and female respondents, although the gap was smaller. The difference in salary for "Director" was approximately $21K or 17.8%, while the difference in salary for "Staff" was approximately $6K or 7.1%. When comparing White to non-White respondents, the gap essentially disappeared for "Director", with a salary difference of less than $2K on average. Interestingly, for "Staff", the relationship flipped: non-White respondents earned $10K more than White respondents on average. This reversal is due to the removal of several non-White interns, consultants and researchers, who collectively had a low average salary of approximately $37.5K.

For every comparison in Tables 3 and 4, we checked if the difference in mean salaries were statistically significant. Our null hypothesis was that the mean salary amongst each of these pairs were equal, and the alternative hypothesis was that the salary of the non-dominant group was lower than that of the dominant group. We performed the Mann Whitney U test and the permutation test with an alpha value of 0.05, and both tests failed to reject the null hypothesis for all comparisons.

*Table 4. The salary distribution in professional sport between directors and staff members, stratified by sex and race.*

|  | Male | Female |  | White | Non-White |
|---|---|---|---|---|---|
| Director (mean ± SD in $,000's) | 140.0 ± 52.4 (n=31) | 118.8 ± 32.5 (n=4) |  | 138.0 ± 51.5 (n=24) | 136.3 ± 49.5 (n=11) |
| Staff (mean ± SD in $,000's) | 87.0 ± 42.9 (n=48) | 81.3 ± 30.0 (n=8) |  | 83.5 ± 38.3 (n=43) | 93.8 ± 48.4 (n=13) |

## 3.2. Experiences with respect to EDI

For the subsequent analysis in this section, as well as the remainder of the paper, we include all 233 respondents, not just those individuals who worked in professional sport.

### 3.2.1. Discrimination

Over one in eight (12.9%) respondents experienced some form of discrimination in the last 12 months (Figure 3). Females experienced discrimination nearly five times more than males (38.2% vs 8.1%). Non-Whites experienced discrimination at a similar proportion to Whites (11.3% vs 11.5%). Conditional on experiencing discrimination, the distributions of the number of discriminatory events, the types of discrimination, and the number of different types of discrimination an individual experienced are shown in Figure 4a), b), and c), respectively. A third of these individuals experienced discrimination five or more times. The most common types of discrimination were sex (23.2%) and racial/ethnic (23.2%). Eleven out of 13 respondents who indicated sexual discrimination were females (one was male and the last respondent did not disclose their sex). Six out of 13 respondents who indicated racial/ethnic discrimination were non-White (six were White and the last respondent did not disclose their race). Our sample consisted of 35 females, 191 males, 69 non-Whites and 162 Whites. Therefore, the percentage of females and non-Whites experiencing sexual and racial/ethnic discrimination (31.4% and 8.7%) was much higher than the corresponding percentage of males and Whites (0.5% and 3.7%), respectively. Respondents who experienced discrimination experienced between one and five types of discrimination, with an average of two.

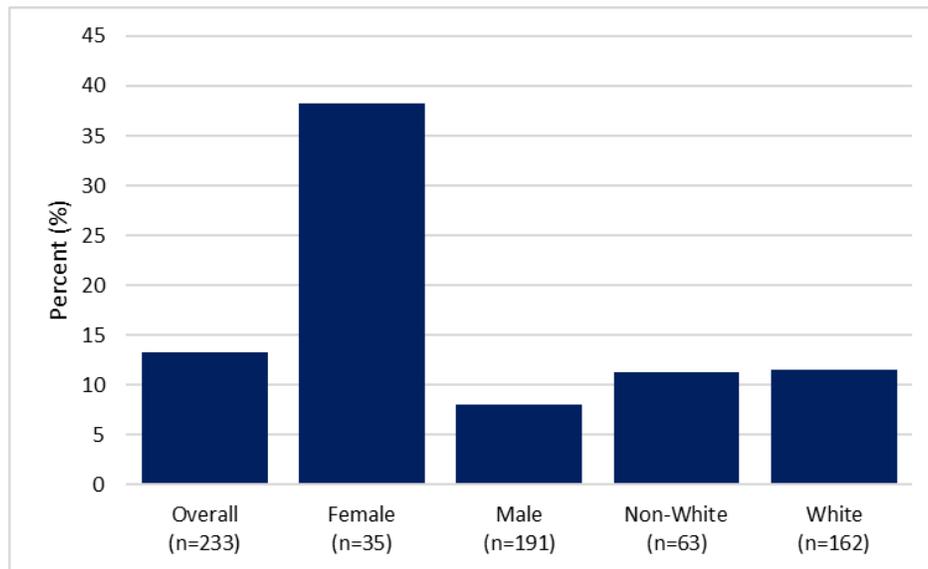

*Figure 3. The percentage of respondents who experienced some form of discrimination in the last 12 months, stratified by sex and race. Note, seven respondents did not indicate their sex, and eight respondents did not include their race.*

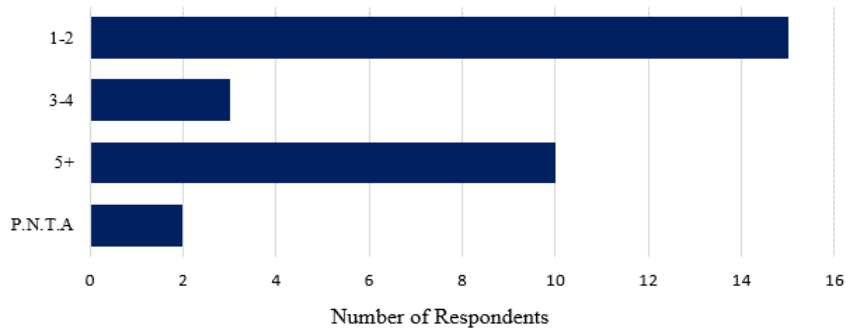
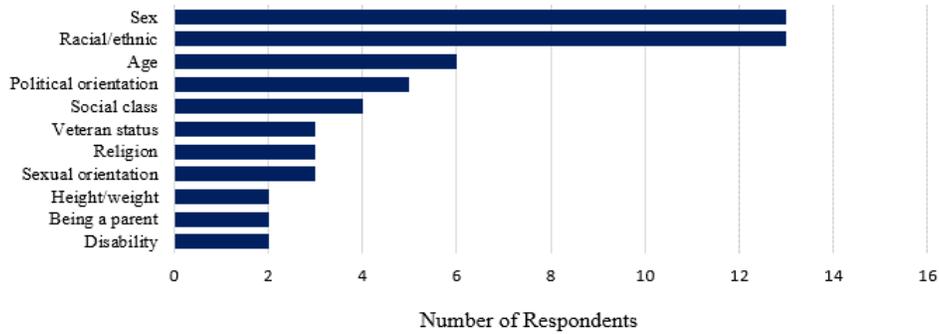
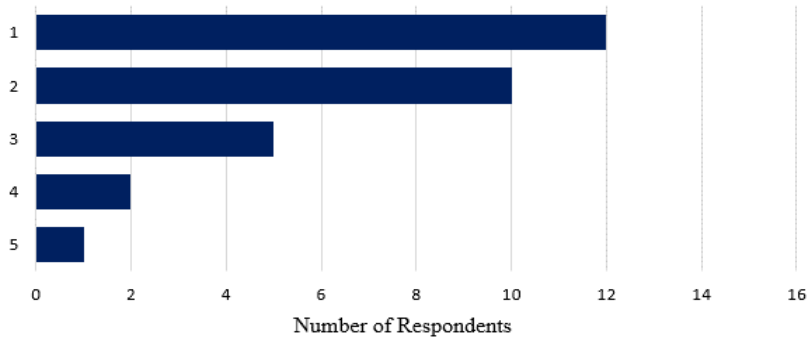

*Figure 4. Breakdown of discriminatory events*

### 3.2.2. Workplace satisfaction

The majority (58.2%) of respondents indicated that they were satisfied or very satisfied with the overall climate of EDI in their workplace. However, there were notable differences when stratifying by sex and race. In particular, females were more than four times as likely to report being "very dissatisfied" (11.4% vs. 2.7%) and 0.6 times as likely to report being "very satisfied" (11.4% vs. 20.5%). In contrast, non-Whites were three times as likely to report being "very dissatisfied" (7.5% vs. 2.5%) but also one and a half times as likely to report being "very satisfied" (26.9% vs. 16.4%). In other words, females were consistently less satisfied by the EDI climate at their workplace, while non-Whites' satisfaction was more polarized. The results below provide more insight into these findings.

**Workplace satisfaction by sex**

Across all the positive workplace descriptors shown in Figure 5, females indicated "strongly agree" far less often than males. For example, 5.9% of females strongly agreed that their workplace was "non-homo/trans/queer phobic" compared to 33.7% of males. Similar but smaller gaps were seen for other descriptors such as "welcoming", "supportive", "non-sexist" and "respectful". Interestingly where males and females responded most similarly was to the descriptor "diverse" – both sexes generally thought that their workplaces were not diverse.

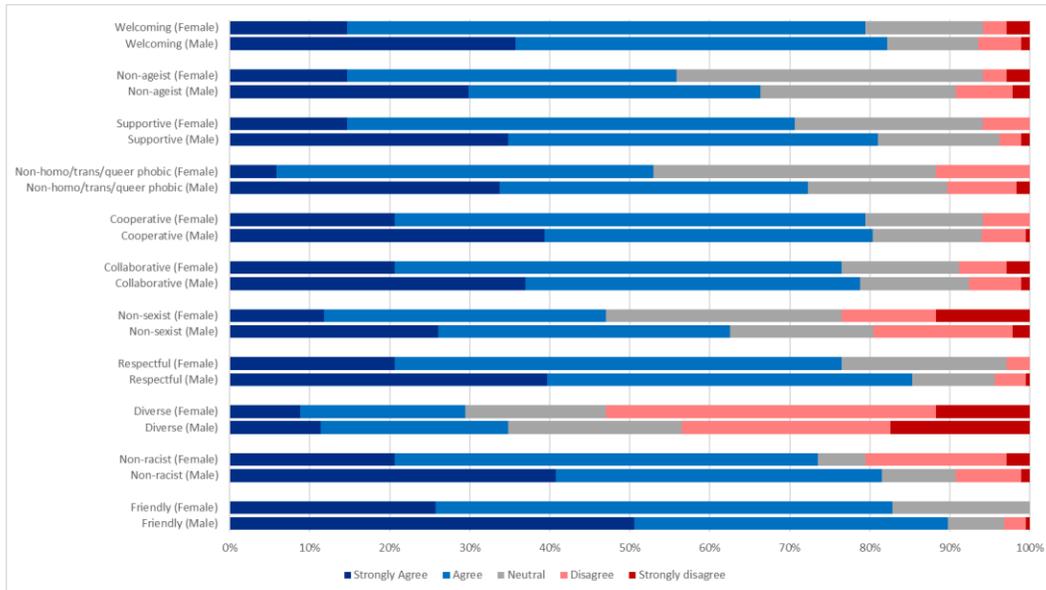

*Figure 5. Descriptors of the workplace stratified by sex.*

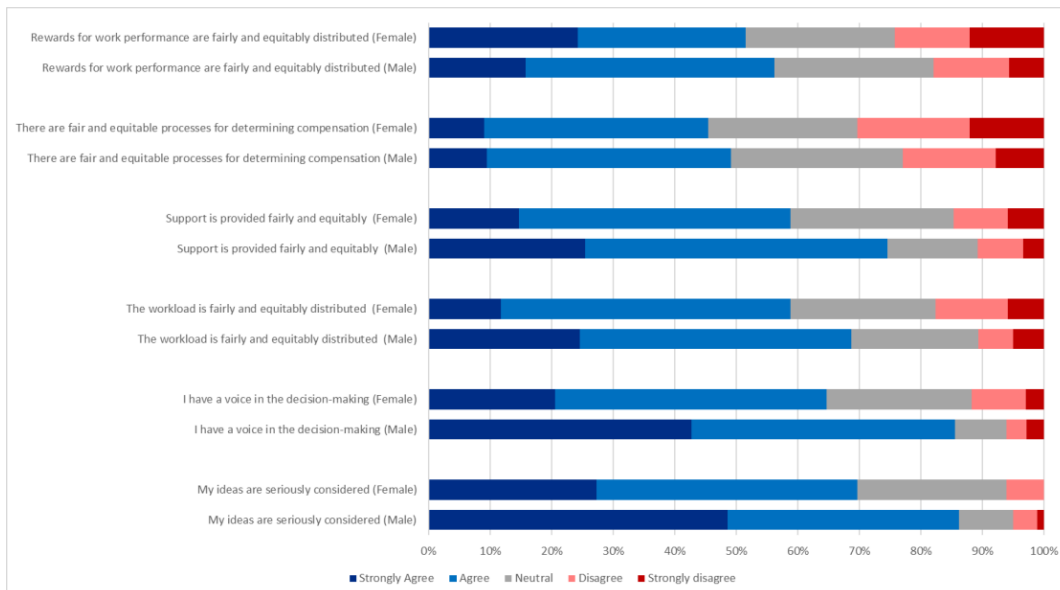

*Figure 6. The experiences of EDI in the respondent's unit/department stratified by sex.*

Figure 6 highlights the respondents' experiences in their particular department, stratified by sex. Interestingly, the first two statements show that males and females have similar levels of agreement regarding fair processes for rewards and compensation. However, females "strongly agreed" less often than males regarding equitable support, equitable workload, having a voice, and being taken seriously.

Figure 7 highlights the respondents' experiences in their specific job or role, stratified by sex. Note that some of these statements were written from a "negative" perspective. Females generally reported more negative experiences in their job. For example, females were twice as likely as males to have considered leaving their job due to isolation or feeling unwelcome (35.3% vs. 17.4%). When restricted to respondents who "strongly agree" with this statement, females were nearly four times as likely as males (14.7% vs. 3.8%). For another example, 36.3% of females strongly agreed that they have to work harder than others to be values equally, compared to 9.8% of males. Males and females answered similarly when it came to questions around fairness of promotions, job performance evaluation and compensation. Looking at all questions from their "positive" perspective, the one that both males and females disagreed with the most is that their employer "provides sufficient programs to foster the success of a diverse staff." Less than 50% of respondents from both sexes agreed with this statement, although females disagreed more strongly. This result complements the finding from Figure 5, where both sexes generally did not view their workplace as diverse.

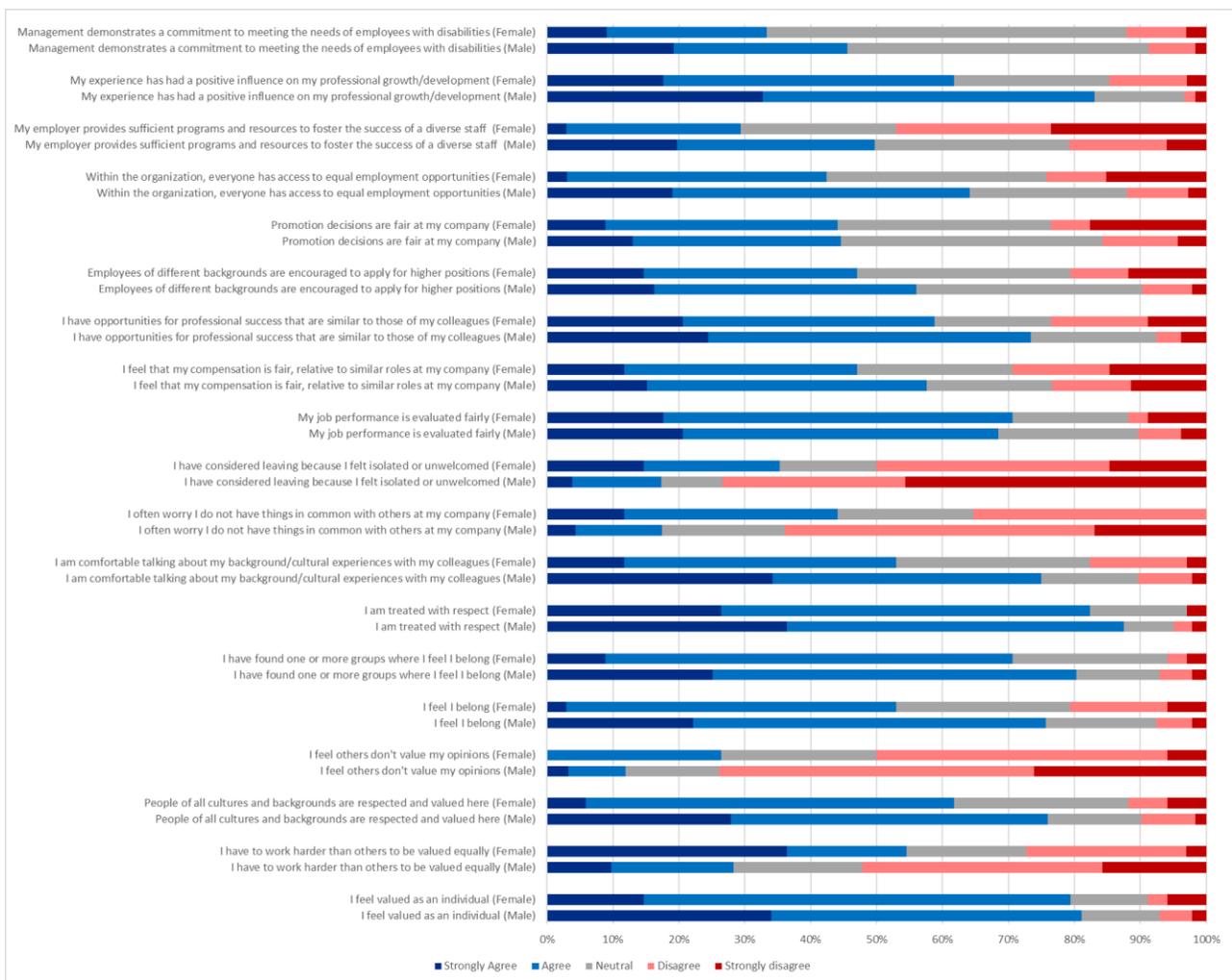

*Figure 7. The experiences of EDI in the respondent's job role stratified by sex.*

## Workplace satisfaction by race

Across all the positive workplace descriptors show in Figure 8, non-White and White respondents shared similar sentiments, with non-Whites being more positive across the board. We find similar trends in Figures 9 and 10, when examining experiences in the respondents' department and specific job, respectively. As an example, consider the statement "Within the organization, everyone has access to equal employment opportunities" in Figure 10. The percentage of non-Whites and Whites who agreed or strongly agreed to this statement was very similar (62.3% vs. 59.4%). Likewise, the percentage of those who disagreed or strongly disagreed was 13.1% vs, 14.8%. The largest difference in responses appeared to be with respect to the workplace descriptor "Diverse" in Figure 8. Over half (51.6%) of non-Whites agreed that their workplace was diverse, compared to 27.1% of Whites.

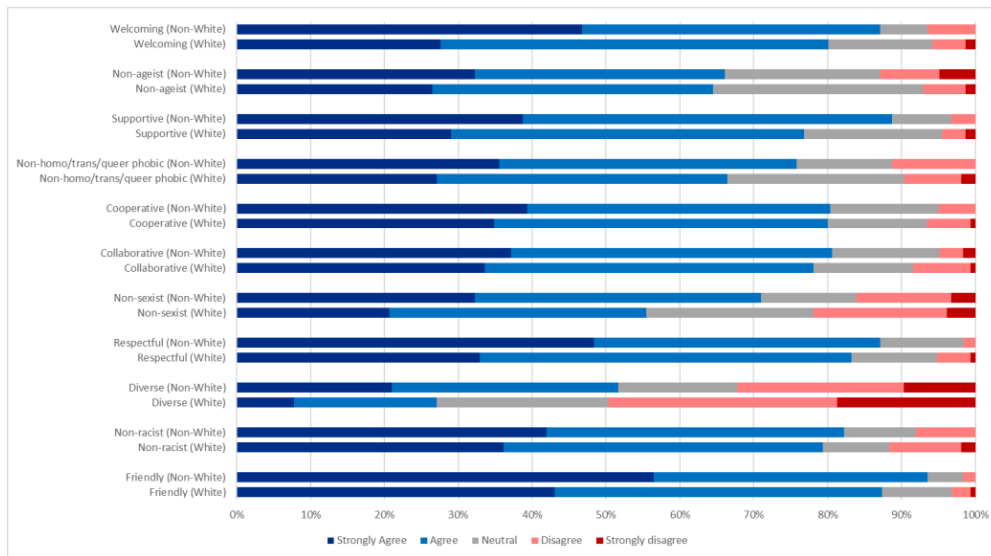

*Figure 8. Descriptors of the workplace stratified by race.*

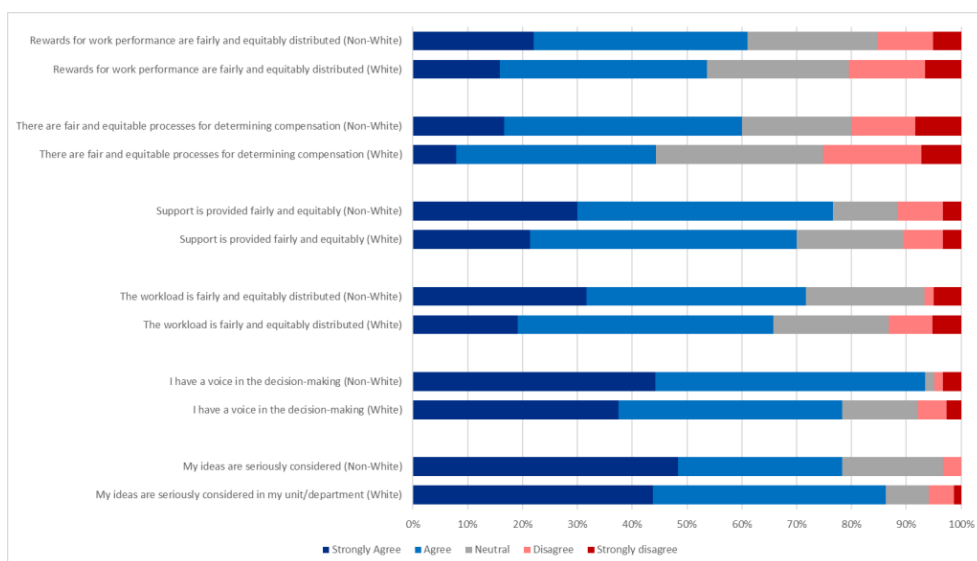

*Figure 9. The experiences of EDI in the respondent's unit/department stratified by race.*

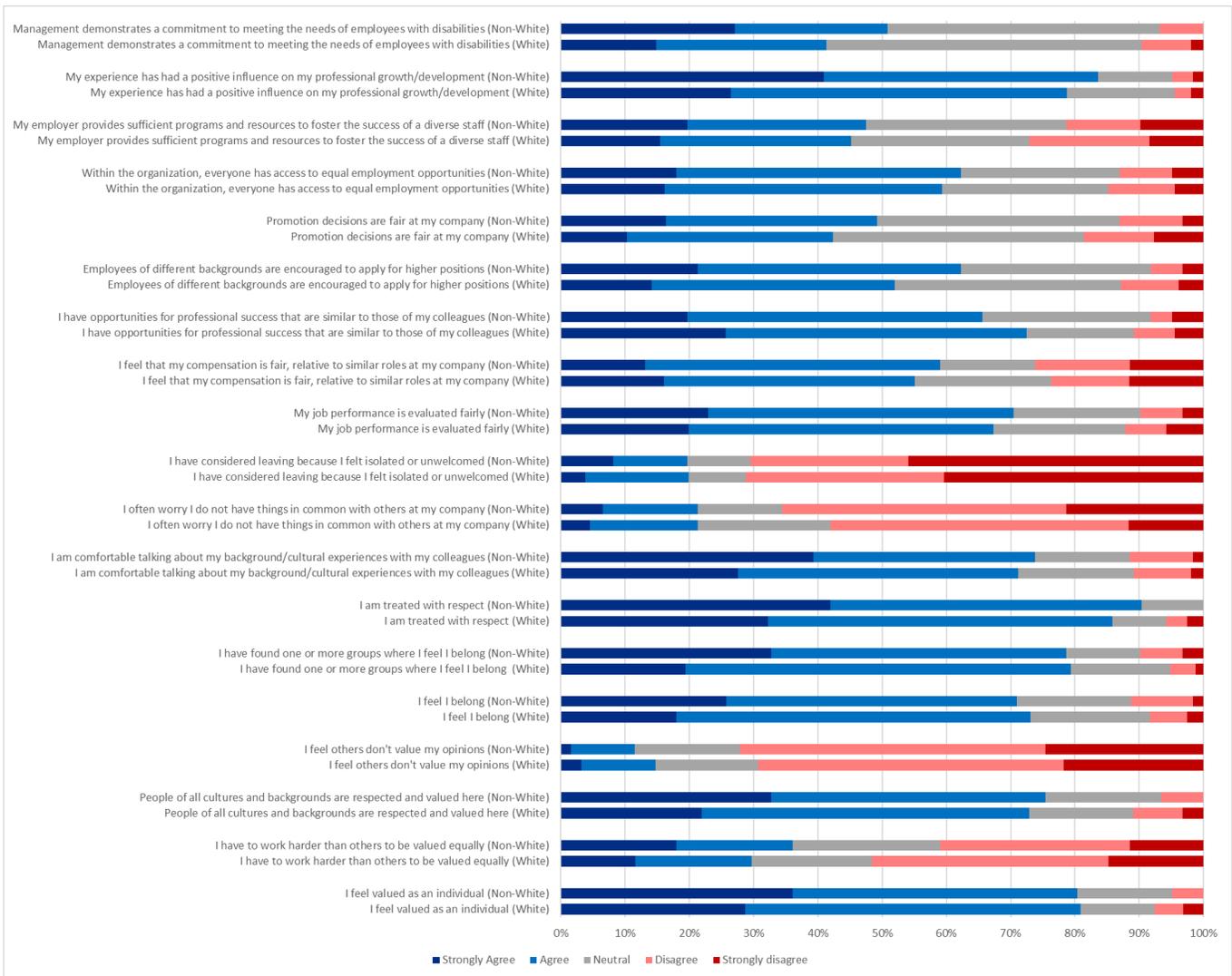

*Figure 10. The experiences of EDI in the respondent's job role stratified by race.*

To better understand the findings in Figures 8-10 (i.e., that non-Whites seem to have slightly higher levels of satisfacation than Whites), we further stratified the respondents by sex and generally found that the results are due to males experiencing higher satisfaction. Note that 89.1% of non-White respondents were male while 82.6% of White respondents were male. In general, both the White and non-White males were generally much more positive about each statement than their female counterparts. As an illustrative example, we broke down the statement "I have a voice in decision making" first by race, and then further by sex (Figure 11). Overall, non-Whites were more positive than Whites in this regard. However, when further stratified by sex, it is the males that have more positive experiences compared to the females in each racial group, and it is this difference in experience that overwhelms any difference due to race. Thus, while it seems that non-Whites have more positive EDI experiences, we believe that this is due to the higher proportion of non-White males than females. We found similar results for the other questions when stratifying by race and sex; we omit them here for brevity.

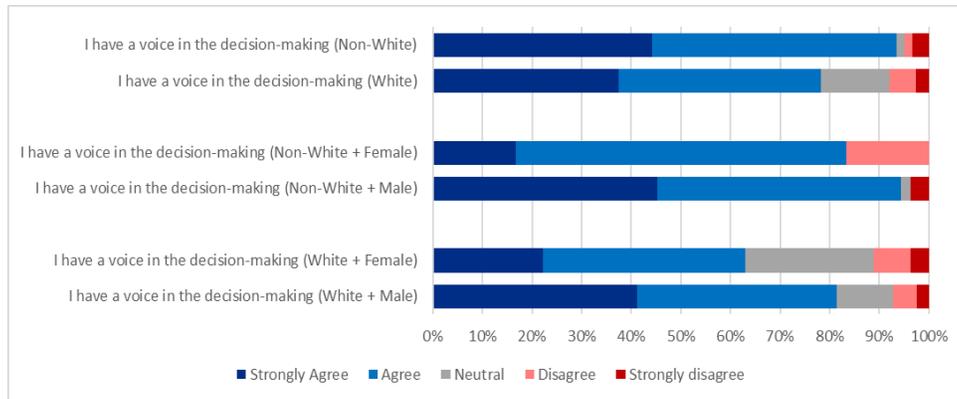

*Figure 11. The responses to the statement "I have a voice in decision-making", stratified by race and sex, both individually and combined.*

## 3.3 Impact of COVID-19

The final two questions in the survey addressed the respondent's workplace experiences since the onset of the COVID-19 pandemic. With many organizations needing to pivot their operations to the virtual environment, the aim was to discover how pandemic-induced changes impacted the respondents. Interestingly, females reported more positive workplace changes, but lower levels of satisfaction, due to the pandemic (Figure 12).

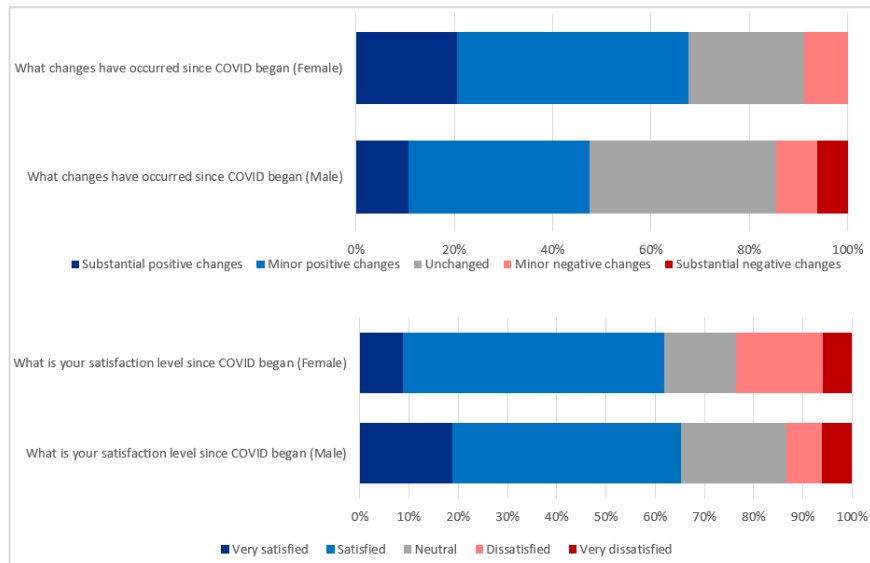

*Figure 12. The changes in the workplace and satisfaction levels since COVID, stratified by sex.*

Figure 13 presents similar results, stratified by race. Both White and non-White respondents similarly reported mostly reported neutral or positive changes due to COVID. However, non-Whites reported higher satisfaction levels compared to Whites. For both Figures 12 and 13, more research and granular questions are needed to determine the reasons for these differences in satisfaction level.

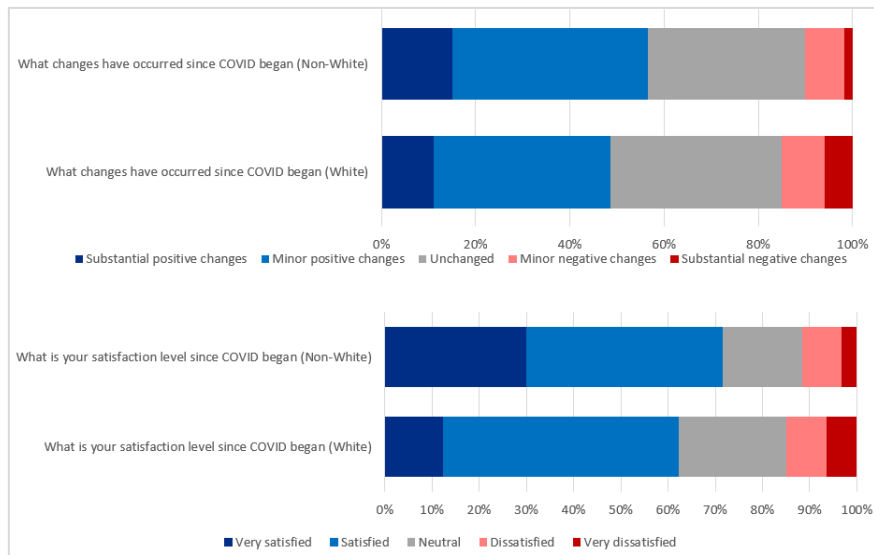

*Figure 13. The changes in the workplace and satisfaction levels since COVID, stratified by race.*

## 4. Discussion

### 4.1. Interpreting the results

Firstly, from the individual demographic variables, we recognize that there are clear dominant groups identified within the sports analytics industry - White, heterosexual, males, young (between the ages of 25-34). The proportions of these groups are heavily skewed, even when compared to other STEM-based industries according to a study by the Pew Research Center (Funk and Parker 2018). For example, across STEM-based jobs, females make up roughly 50% of the population, compared to only 15% of respondents in our sports analytics survey. Similarly, Black employees hold roughly 9% of STEM jobs, compared to only 1.7% in our sample. Thus, our survey suggests that sports analytics is critically lacking in diversity, even in comparison to STEM in general.

In contrast, when it comes to advanced education degree holders, our study suggests that people working in sports analytics had much higher representation compared to general STEM-based jobs. Indeed, across STEM-based jobs, approximately 29% of workers hold a postgraduate degree (Funk and Parker 2018). In our sample, it is nearly double that percentage (57.5%). Although diversity within STEM education has been steadily increasing, there is a still a gap when it comes to racialized or marginalized communities pursuing advanced degrees, which may help explain the above findings. Within respondents who held a doctorate, 37.0% worked in a traditional academic organization and half of them (50.0%) worked for professional sports or sports tech companies. Within the advanced degree holders in our survey, there is a much greater prevalence of technical degrees such as statistics, computer science, math and engineering (e.g., 73.9% of doctorate degrees) compared to more traditional sport-related degrees, such as kinesiology or sport medicine (e.g., 15.2% of doctorate degrees). This reinforces the perception of sports analytics as a field that favors analytics-based personnel with specialized and focused technical skills, over individuals trained in the sport sciences or business/management.

Within professional sport, our data suggests that there are clear imbalances in representation by sex across different sports analytics job levels, with a much higher proportion of males holding senior positions. Indeed, within our survey the only job level in which there is parity between sexes is at the intern level. Studies of the workforce at large find a similar result (Burns et al. 2021), namely, that as positions become more senior, there is less female representation. Our survey responses also indicate a salary gap between male and female respondents. A 2020 study by the Pew Research Center shows that females earn 84% of what males earn when analyzing median salaries of full-time and part-time workers in the United States ([Barrosa and Brown 2021](#)). Our results were similar. In our sample, females in management-level jobs earned 78% of what males earned, while females in non-management jobs earned 82% of what males earned. Similar gaps in representation and salary were observed, with Whites having higher representation and higher salaries, though the gaps were less pronounced. We acknowledge that none of the differences in mean salaries were statistically significant, however, we believe that these tests were likely underpowered given the few female and non-white respondents – especially in senior roles.

Within our analysis of discrimination, we observed that an equal number of White and non-White respondents experienced racial discrimination. However, this translated to 3.7% and 8.7% of White and non-White respondents, respectively. We acknowledge that within the literature, it has been shown that these two groups experience racial discrimination differently. Due to the smaller and heterogenous sample of non-White respondents, we cannot make any generalizations from this result. Moreover, the survey relied on each respondent to answer questions regarding discrimination with their own definitions of what constituted as discriminatory. We hope future studies can provide an education or explanatory component to help all respondents arrive at the same definitions of these terms.

We also comment on our finding that when an individual experienced discrimination, they received anywhere between one to five unique types of discrimination, with an average of two. This highlights an important notion of intersectionality. Intersectionality is a way of explaining complexities in the world and people by viewing characteristics of an individual (i.e., race, class, gender, etc.) as interrelated and mutually shaping one another ([Collins and Bilge 2020](#)). Essentially, inequities are not the result of a single factor, but rather the outcome of multiple different factors that are not simply additive in nature. The presence of individuals experiencing multiple forms of discrimination provide evidence of the existence of intersectionality within our study sample. We hope future studies can more explicitly gather insights into how intersectionality is experienced by those in the sports analytics industry.

In terms of experiences with respect to EDI, there are clear positive advantages afforded to males within the sports analytics industry that females do not receive  For example, females were nearly five times as likely to experience discrimination compared to males (38.2% vs. 8.1%). A recent study across all STEM-based jobs reported a similar finding, but focused on gender, instead of sex: 50% of women and 19% of men experienced gender discrimination ([Funk and Parker 2018](#)). In general, females reported lower workplace satisfaction than males, for example with respect to having a voice in decision-making, having to work harder to be valued equally, and working in a welcoming and supportive workplace. Interestingly, despite the fact that females earned less than males in our sample, they reported similar levels of satisfaction with respect to fair and equitable processes for determining compensation.

In contrast to the stratification by sex, the responses about workplace EDI experience when stratified by race were quite similar. Our results suggest that non-Whites had slightly more positive experiences than their White colleagues, however, this may be due to sample noise. This finding is counter to the general literature, which demonstrates clear evidence that racialized communities have more negative experiences in the workplace ([Smith and Calasanti 2005](#)). The further stratification of our results by sex may explain the discrepancy: the proportion of males in the non-White group was 6.5% higher than in the White group, and it was the much higher satisfaction of males compared to females across all racial groups that accounted for the slightly more positive experience of the non-White group. Another issue is that we have a relatively small and imbalanced dataset. In our sample, White respondents outweighed non-White respondents by a 2.3:1 ratio. We also note that comparisons with respect to race found in the literature are primarily done between White individuals and individuals from a specific racialized community (i.e., Black, Latinx, Indigenous, etc.). We acknowledge that this analysis may be challenging to conduct if there are relatively few racialized persons currently employed in sports analytics. Lastly, our sample of non-White respondents is very heterogeneous, which renders any generalizations about this group's experiences difficult and subject to added noise. Overall, more research is needed to understand if there is a clear difference in EDI experience across racial communities in sports analytics.

We also note that some changes occurred as a result of the COVID-19 pandemic. For example, females reported more positive workplace changes, but lower satisfaction levels compared to males. We cannot conclusively comment on what is driving this trend, but we conjecture that although widespread work-from-home practices provided greater flexibility (a positive change), it ultimately led to greater amounts of office-work and housework for females, leading to more burnout and lower satisfaction (Dunatchik et al. 2021, Burns et al. 2021, Lyttelton et al. 2021).

## 4.2. Limitations

This study has several limitations. While we made every effort to share our survey as widely as possible, we could not confirm which particular sports, teams or personnel received notice of the study. We suspect the sports analytics population is much larger than our survey sample, and it is unclear if the respondents are representative of the overall population. We hope the preliminary results presented in this paper lead to additional, larger-scale studies of EDI in the field. In particular, future studies should include more qualitative data collection to obtain richer details on individuals' EDI experiences.

Many of our analyses were affected by small sample sizes. Results for groups with small samples were either omitted or are limited in their robustness. It is unclear if these small numbers are due to a low response rate or point to the fact that these groups are under-represented in the sports analytics population. Moreover, the comparison of responses between groups may also be affected by the underlying differences between the groups, rather than a true difference in their responses. Salary comparisons were restricted to means, rather than medians, because our large salary buckets coupled with a small sample would have made comparisons challenging.

We also acknowledge that completion of the survey was voluntary, and as such, may not have attracted a representative sample. For example, the survey may be biased with respondents who simply had the time to fill out the survey. Alternatively, it may have attracted respondents who felt more strongly about EDI issues than the average individual. Moreover, respondents were expected to interpret the phrasing and terminology of the questions based on their own judgement. Given that the definitions around equity, diversity and inclusion are continuing to be defined, this may have impacted some of the survey results.

## 5. Recommendations

We conclude with some recommendations that are based on our specific findings with respect to EDI in sports analytics.

1. **Diversity-promoting initiatives.** Our data suggests that continued effort is needed to increase participation of underrepresented communities in sports analytics and to foster their success. Possibilities include: targeted scholarships for individuals from underrepresented communities doing research in sports analytics in a university setting; a conscious focus on including invited conference speakers and panelists from diverse communities; training and mentorship programs for underrepresented racialized or marginalized communities that address their unique needs and barriers.

2. **Salary transparency and female support.** Females believe they are treated fairly with respect to compensation, despite the fact that they have lower salaries. Improved salary transparency can provide females with useful data to negotiate for more equitable pay. Additionally, recruitment and job-development pipelines that support females in the industry may help close the current pay gap.

3. **Create an EDI office.** Organizations are increasingly hiring staff dedicated to EDI. This office can be responsible for education of the organization's employees, ensuring that underrepresented individuals receive appropriate mentorship, and advocating for accountability and transparency when it comes to negative workplace experiences. In particular, there could be a dedicated officer to whom episodes of discrimination would be reported, or who would advocate for equitable workloads and shared decision making. Executive level support would be needed to endorse and implement these policies and practices.

4. **Commitment to greater research.** Larger studies are needed to avoid the small sample size issues encountered in this paper. Future studies should also include qualitative data and long-format interviews. These broader and more detailed studies would allow greater clarity and insight into the experiences of members of racialized or marginalized communities, especially if sample sizes remain smaller.

# 6. Acknowledgements

This project was funded by a Connaught Global Challenge Award from the University of Toronto. The authors thank Caroline Fusco and Robin Waley for their expert recommendations during survey development. The authors are grateful to Farah Bastien, David Clarke, Nicole DeFord, Doug Fearing, Jessica Gelman, Graham Goldbeck, Shayna Goldman, Valerie La Traverse, Richard Lapchick, Mike Lopez, Andrew Maker, Shannon Osborne, Seth Partnow, Devin Pleuler, Sarah Steed, Ming-Chang Tsai, Samuel Ventura, Seth Walder, and Darin White for their help with outreach and spreading the word about the survey during the course of this project. Finally, we are grateful to Stephanie Kovalchik and Keith Willoughby for helpful feedback on earlier drafts of the paper.

# Appendix

The following section contains all of the survey questions.

### **Section 1 - Current job**

1. Consent

2. Are you responsible for performing data analytics regularly (i.e., multiple times per week) in your current job?
   Yes or No

3. Are you responsible for interpreting analyses regularly (i.e., multiple times per week) in your current job?
   Yes or No

4. What type of organization is your current (primary) job?
   - Professional sport (e.g., NBA, NFL, MLB, etc.)
   - Elite Amateur sports (e.g., Olympic/Paralympic team or National Sport Organization)
   - Collegiate sports
   - Sports/fitness technology company
   - Community sport organization (e.g., club)
   - Sport Entertainment
   - Academic/research institution
   - Other (text field)

5. Please select the option that best describes the department/division of your current (primary) job?
   - Sport Performance
   - Sport Medicine
   - Coaching Staff
   - Business Development
   - Business Operations
   - Technology/Digital
   - Marketing
   - Analytics
   - Computer science
   - Statistics/Mathematics
   - Engineering
   - Kinesiology/Sport Science
   - Other (text field)

6. Select the option that best describes the level of your current (primary) job.
   - Vice president
   - Director (e.g., of a department)
   - Manager/supervisor (e.g., of specific disciplinary group)
   - Professional staff member (e.g., sports coach, data analyst, sport scientist, S&C coach)
   - Faculty (e.g., asst., assoc., full professor)
   - Researcher
   - Student
   - Intern
   - Other (text field)

7. How long have you worked in your current (primary) job?
   - Less than one year
   - 1–2 years
   - 2–4 years
   - 4–6 years
   - More than 6 years

8. What is the income (in USD) for your current (primary) job?
   - Volunteer
   - Less than $25,000
   - $25,000 to $49,999
   - $50,000 to $74,999
   - $75,000 to $99,999
   - $100,000 to $124,999
   - $125,000 to $149,999
   - $150,000 to $174,999
   - $175,000 to $199,999
   - More than $200,000
   - Prefer not to answer

**Section 2 - General (demographic) Information**

9. What is your age?
   - 19 years old and younger
   - 20-24 years old
   - 25-29 years old
   - 30-34 years old
   - 35-39 years old
   - 40-44 years old
   - 45-49 years old
   - 50 years and older

10. Please select the biological sex you were assigned at birth (listed on your original birth certificate).
    - Male
    - Female
    - Undetermined/Intersex

- Prefer not to answer

11. Please select the option that best describes your current gender identity.
    - Man/Primarily masculine
    - Woman/Primarily feminine
    - Neither man/masculine or woman/feminine (e.g., gender non-conforming, genderqueer, non-binary)
    - Transgender man
    - Transgender woman
    - Other cultural gender minority (e.g., two-spirit)
    - Prefer not to answer
    - Another gender (text field)

12. Please select the option that best describes your current sexual orientation.
    - Heterosexual (straight)
    - Bisexual
    - Gay
    - Lesbian
    - Queer
    - Asexual
    - Prefer not to answer
    - Other (text field)

13. Please select the option that best describes your relationship status.
    - Single
    - Committed relationship (not married)
    - Married/domestic partnership
    - Widowed
    - Divorced
    - Separated
    - Prefer not to answer
    - Other (text field)

14. Please select the categories that best describe your race. Select all that apply.
    - Black (African, Afro-Caribbean decent)
    - East/Southeast Asian (Chinese, Korean, Japanese, Taiwanese, Filipino, Vietnamese, Thai decent)
    - Indigenous (Native American, First Nations, Métis, Inuit decent)
    - Latinx (Latin American, Hispanic decent)
    - Middle Eastern (Arab, Persian, West Asian decent)
    - South Asian (India, Pakistani Sri Lankan, Bangladeshi decent)
    - White (European decent)
    - Prefer not to answer
    - Other (text field)

15. Please select the options that best describe your religion and/or spiritual affiliation? Select all that apply.
    - Buddhist
    - Christian
    - Hindu
    - Jewish
    - Muslim
    - Sikh
    - Indigenous Spirituality
    - No religion
    - Prefer not to answer
    - Other (text field)

16. If you identify as a person who has a disability, then please select all that apply.
    - I do not identify as a person with a disability
    - Vision
    - Hearing
    - Speech/verbal communication
    - Mobility/Flexibility/Dexterity
    - Pain
    - Learning
    - Developmental
    - Memory
    - Mental health-related
    - Prefer not to answer
    - Other (text field)

17. What is the highest degree you have earned?
    - High school diploma or GED
    - Associate's degree (Two-year college)
    - Bachelor's degree (Four-year college/university)
    - Master's degree (e.g., MA, MS, MBA, MPH)
    - Doctorate (e.g., Ph.D., Ed.D.)
    - Professional degree (e.g., MD, DO, DDS, DVM)
    - Law degree (e.g., JD, LLM, SJD)
    - Other degree (text field)

18. What field/discipline is the highest degree you have earned?
    - Business management
    - Marketing
    - Computer Science
    - Statistics/Mathematics
    - Engineering
    - Kinesiology/Sport Science
    - Sports Medicine
    - Other (text field)

## Section 3 - Equity and Inclusion

19. In general, how satisfied or dissatisfied are you with the climate/environment that you have experienced in your current (primary) job in the 12 months prior to COVID, specifically related to equity, diversity and inclusion?

    - Very Satisfied
    - Satisfied
    - Neutral
    - Dissatisfied
    - Very Dissatisfied

20. How you would rate your current (primary) work environment/experiences on the following items (prior to COVID) (response options: Strongly Agree – Agree – Neutral – Disagree – Strongly Disagree):
    - Friendly
    - Non-racist
    - Diverse
    - Respectful
    - Non-sexist
    - Collaborative
    - Cooperative
    - Non-homo/trans/queer-phobic
    - Supportive
    - Non-ageist
    - Welcoming

21. Considering your experiences over the 12 months prior to COVID, please indicate your level of agreement with each of the following statements for your current (primary) job (response options: Strongly Agree – Agree – Neutral – Disagree – Strongly Disagree):
    - I feel valued as an individual.
    - I have to work harder than others to be valued equally.
    - People of all cultures and backgrounds are respected and valued here.
    - I feel others don't value my opinions.
    - I feel I belong.
    - I have found one or more groups where I feel I belong.
    - I am treated with respect.
    - I am comfortable talking about my background/cultural experiences with my colleagues.
    - I often worry I do not have things in common with others at my company.
    - I have considered leaving because I felt isolated or unwelcomed.
    - My job performance is evaluated fairly.
    - I feel that my compensation is fair, relative to similar roles at my company.
    - I have opportunities for professional success that are similar to those of my colleagues.
    - Employees of different backgrounds are encouraged to apply for higher positions.
    - Promotion decisions are fair at my company.
    - Within the organization, everyone has access to equal employment opportunities.
    - My employer provides sufficient programs and resources to foster the success of a diverse staff.
    - My experience has had a positive influence on my professional growth/development.

- Management demonstrates a commitment to meeting the needs of employees with disabilities.

22. During the 12 months prior to COVID, did YOU personally experience any discriminatory events at your current (primary) job?
    - Yes
    - No
    - Prefer not to answer

23. If Yes to #22: How often did it occur within those 12 months?
    - 1-2 times
    - 3-4 times
    - More than 4 times
    - Prefer not to answer

24. If yes to #22: What was the discriminatory event related to? Select all that apply.
    - Ability or disability status
    - Racial or ethnic identity
    - Sex
    - Sexual orientation
    - Gender identity or gender expression
    - Veteran status
    - Marital status
    - Age
    - Religion
    - Height or weight
    - Political orientation
    - Social class
    - Other

25. Considering your experiences over the 12 months prior to COVID, please indicate your level of agreement with each of the following statements for your "primary work unit/department."
    (response options: Strongly Agree – Agree – Neutral – Disagree – Strongly Disagree):
    - My ideas are seriously considered in my unit/department.
    - I have a voice in the decision-making that affects my work in my unit/department.
    - The workload is fairly and equitably distributed in my unit/department.
    - There are fair and equitable processes for determining compensation in my unit/department.
    - Support is provided fairly and equitably in my unit/department.
    - Rewards for work performance are fairly and equitably distributed in my unit/department.

26. In general, what types of changes have occurred in your current (primary) work environment since COVID began (approximately March 2020) specifically related to equity, diversity and inclusion?

    - Substantial positive changes
    - Minor positive changes
    - Unchanged
    - Minor negative changes
    - Substantial negative changes

27. In general, how satisfied or dissatisfied are you with the climate/environment that you have experienced in your current (primary) job since COVID began (approximately March 2020), specifically related to equity, diversity and inclusion?

- Very Satisfied
- Satisfied
- Neutral
- Dissatisfied
- Very Dissatisfied